# Scaling Properties of Ge-Si$_x$Ge$_{1-x}$ Core-Shell Nanowire Field Effect Transistors

Junghyo Nah, *Student Member*, *IEEE*, En-Shao Liu, Kamran M. Varahramyan, Davood Shahrjerdi, Sanjay K. Banerjee, *Fellow, IEEE,* and Emanuel Tutuc, *Member, IEEE*

*Abstract*— We demonstrate the fabrication of high-performance Ge-Si$_x$Ge$_{1-x}$ core-shell nanowire field-effect transistors with highly doped source and drain, and systematically investigate their scaling properties. Highly doped source and drain regions are realized by low energy boron implantation, which enables efficient carrier injection with a contact resistance much lower than the nanowire resistance. We extract key device parameters, such as intrinsic channel resistance, carrier mobility, effective channel length, and external contact resistance, as well as benchmark the device switching speed and ON/OFF current ratio.

## I. INTRODUCTION

RECENT years have witnessed a remarkable progress in emerging research materials, such as semiconductor nanowires or carbon nanotubes, as alternatives to conventional complementary metal-oxide-semiconductor (CMOS) technology [1-5]. A key question regarding such devices, relevant to both benchmarking potential application as well as gaining insight into fundamental electronic properties, is the device performance scaling with channel length. For carbon nanotubes it has been experimentally established that the nanotube resistance scales linearly with length for channel lengths larger than a few microns, where diffusive transport applies, and is independent of length for channel lengths smaller than one micron, in the ballistic transport regime [4]. Here we present the first scaling study of high performance germanium (Ge) – silicon-germanium (Si$_x$Ge$_{1-x}$) core-shell nanowire (NW) field-effect transistors (FETs) with highly doped source (S) and drain (D). The highly doped ($>10^{20}$ cm$^{-3}$) source and drain, realized using boron (B) ion implantation, enable efficient carrier injection with a contact resistance much lower than the nanowire resistance. The nanowire FET resistance scales linearly with the channel length down to 300 nm, indicating that the transport in these nanowires is diffusive at room temperature.

Semiconductor nanowires enable the realization of novel device geometries, such as gate-all-around field effect transistors, which allow for more energy efficient electronics at a given switching speed, thanks to a better electrostatic control of the channel [6-9]. Germanium and Ge-Si core-shell nanowires have attracted interest as a platform for aggressively scaled field effect transistors, thanks to Ge higher carrier mobility than Si, and its compatibility with CMOS technology [10-12]. A main, albeit mundane obstacle, that has often impeded both an accurate electrical characterization, as well as the realization of high performance devices using nanomaterials is the carrier injection. Generally NW FETs employ metal contacts at the source and drain terminals [13-14], which limits the device performance because of the Schottky barrier existent at the metal-semiconductor interface. Moreover, ambipolar behavior is usually observed in such devices. The contact material that provides low contact resistance and unipolar carrier injection should be highly conductive, with a Fermi level aligned with the nanowire conduction or valence band, depending on the carrier type to be injected. A highly doped section of the same nanowire satisfies these conditions. Nanowire doping with axial modulation can be achieved via vapor-solid-liquid (VLS) growth mechanism [15], thermal diffusion from dopant-containing molecule [16], and ion implantation [17-20]. Ion implantation allows for accurate axial doping control along the NW and is widely used in existing CMOS technology. Here, we employ low energy ion implantation in order to realize NW FETs with highly doped source and drain.

## II. FABRICATION OF GE-SI$_x$GE$_{1-x}$ CORE-SHELL NW FETS

Our samples consist of Ge-Si$_x$Ge$_{1-x}$ epitaxial core-shell nanowires. The core-shell NWs were grown in an ultra-high-vacuum (UHV) chemical vapor deposition (CVD) chamber, via VLS mechanism and using Au as a catalyst. First, the Ge core was grown at a total pressure of 5 Torr and a wafer temperature of 285 °C using 60 sccm GeH$_4$ (10 % diluted in He). Next, an *epitaxial* Si$_x$Ge$_{1-x}$ shell was grown in UHV conditions in the *same* chamber, by co-flowing 7 sccm SiH$_4$ and 60 sccm of GeH$_4$ at a wafer temperature of 400 °C. Using transmission electron microscopy coupled with energy dispersive X-ray spectroscopy, we deduce the Si$_x$Ge$_{1-x}$ shell thickness of ~4 nm and a Si content in the shell of $x$ = 0.3 [Figure 1b]. The role of the Si$_x$Ge$_{1-x}$ shell is two-fold. First, it acts as a passivation layer for the Ge surface, which is known to have a high density of interface traps in contact with a

Manuscript received July 23, 2009. This work was funded by DARPA contracts HR0011-08-1-0050 and N66001-07-12013. Junghyo Nah, E. -S. Liu, K. M. Varahramyan, D. Shahrjerdi, S. K. Banerjee, and E. Tutuc are with the Microelectronics Research Center, University of Texas, Austin, TX 78758 USA (e-mail: jnah@ieee.org, etutuc@mer.utexas.edu).

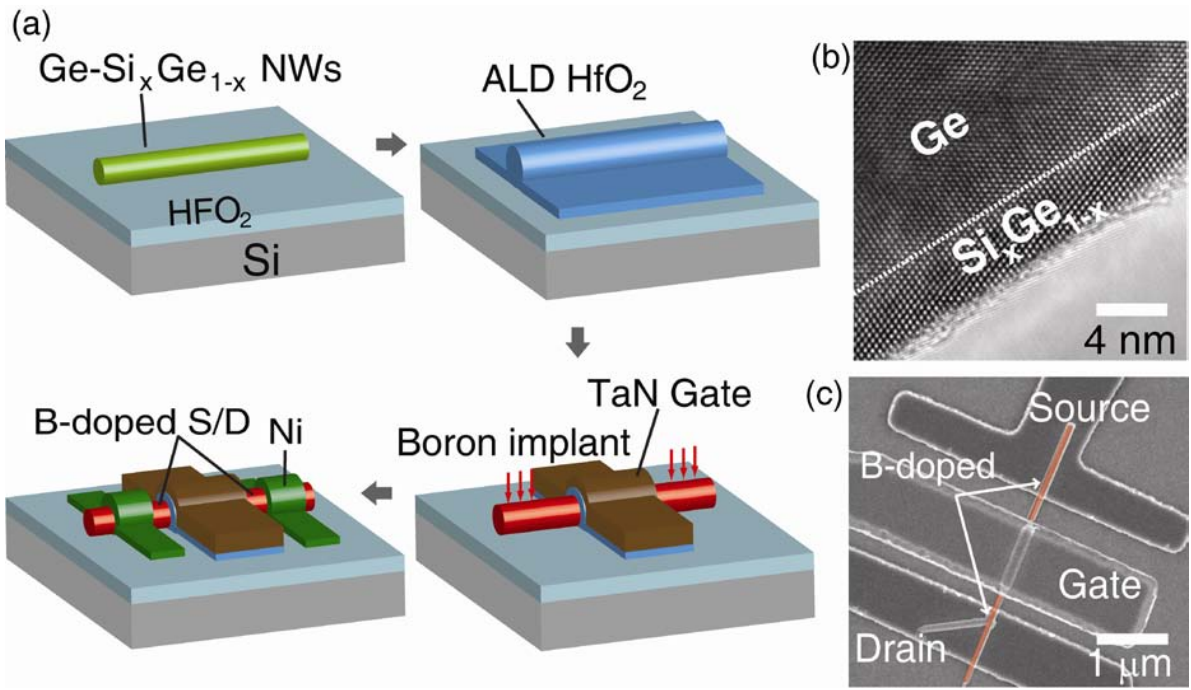

**Figure 1. Top-gated Ge-Si$_x$Ge$_{1-x}$ NW FET. a.** Schematic representation of a top-gated Ge-Si$_x$Ge$_{1-x}$ NW FET and fabrication flow. **b.** Transmission electron micrograph of a Ge-Si$_x$Ge$_{1-x}$ core-shell NW, evincing a single crystal shell grown epitaxially on the Ge core. **c.** Scanning electron micrograph showing a top-gated NW FET device. The red regions represent the highly boron-doped NW sections.

dielectric, and enables the realization of inversion layers in germanium. Secondly, thanks to a positive band offset between Si$_x$Ge$_{1-x}$ and Ge valence band, it serves as a barrier and confines the holes in the Ge core. The UHV CVD in-situ shell growth allows for the Si$_x$Ge$_{1-x}$ shell thickness and content to be engineered with minimum impurity incorporation. In particular, here we chose a reduced Si content, x = 0.3, in order to minimize the strain in the Ge-Si$_x$Ge$_{1-x}$ core-shell heterostructure, while still maintaining the interface passivation and hole confinement noted above.

Figure 1 shows a schematic of the top-gated NW FET with highly doped source and drain [Figure 1(a)], along with a transmission electron microscopy of Ge-Si$_x$Ge$_{1-x}$ NW [Figure 1(b)], and a scanning electron micrograph of the device [Figure 1(c)]. The fabrication process flow is briefly described in the following. Post growth the Ge-Si$_x$Ge$_{1-x}$ core-shell NWs were suspended in ethanol and dispersed onto a HfO$_2$ (10 nm)/Si (100 *n*-type) substrate. The wafer with dispersed NWs was then cleaned with a 2 % hydrofluoric acid (HF) solution for 20s and deionized (DI) water for 20s for 2 cycles, before the gate oxide deposition. Next, a 9 nm-thick HfO$_2$ layer was deposited by atomic layer deposition (ALD) at 250 °C. The equivalent oxide thickness (EOT) of deposited gate oxide was ~3.9 nm, evinced by capacitance-voltage measurement on planar capacitors processed in parallel with the device. The gate electrode was defined by e-beam lithography (EBL), followed by 120 nm of TaN (tantalum nitride) deposition and lift-off. In order to remove resist residues, the device was cleaned with O$_2$ plasma for 10s (50 W). The HfO$_2$ layer deposited on the S/D areas of the device was etched by diluted HF (~3 %). Once the NW FET gate areas are defined, the samples are ion-implanted with boron which results in highly doped NW areas *outside* the TaN metal gate. The relatively thick TaN metal gate prevents the NW FET channel from B penetration. The B ion implantation was done at an ion energy of 3 keV, with a dose of $10^{15}$ cm$^{-2}$, and rotating 360° with 32° tilt during the ion implantation. The devices then underwent a rapid thermal anneal at 600 °C 5min in an N$_2$ ambient to activate the implanted dopants. We expect that implant-induced crystal damage in NWs be removed after activation anneal thanks to the Ge's faster defect removal and regrowth velocity compared to Si [21]. Subsequently, the S/D contacts were defined by EBL, metal (Ni) deposition and lift-off. A one minute anneal at 300 °C completes the NW FET fabrication. Based on a systematic study of the electronic properties (doping concentration and mobility) of B-implanted Ge-Si$_x$Ge$_{1-x}$ core-shell NWs, we expect a doping concentration of $10^{19}$ ~ $10^{20}$ cm$^{-3}$ in the B ion implanted sections of the NW, a NW resistivity of $2.6 \times 10^{-3} \pm 1.9 \times 10^{-4}$ Ω·cm, and a Ni-NW specific contact resistivity of $1.1 \times 10^{-9} \pm 2.2 \times 10^{-10}$ Ω·cm$^2$, corresponding to contact resistances of $300 \pm 200$ Ω [19]. This step is the key to enable efficient, unipolar hole injection in the NW FETs, as well as low external contact resistance. To probe the scaling properties of Ge-Si$_x$Ge$_{1-x}$ NW FETs, we fabricated devices with different channel lengths, ranging from 300 nm to 1 μm.

### III. RESULTS AND DISCUSSION

In order to characterize the devices, we measure either the drain current ($I_d$) as a function of drain bias ($V_d$) at constant gate bias ($V_g$) (output characteristics), as well as $I_d$ vs. $V_g$ at constant $V_d$ values (transfer characteristics). Figure 2 shows $I_d$ vs. $V_d$ and $I_d$ vs. $V_g$ data, measured for several Ge-Si$_x$Ge$_{1-x}$ NW FETs with different channel lengths ($L_g$), from $L_g$ = 300 nm to $L_g$ = 1 µm. The NW diameters in these devices are similar, $d$ = 52 ± 4 nm. The drain current normalized to the NW diameter ($d$), namely the output current per footprint, are shown on right axis of the $I_d$ - $V_d$ graphs to facilitate a comparison of the device characteristics. Two observations are apparent from the Figure 2 data. First, the device characteristics clearly show unipolar behavior, by comparison to Schottky metal-semiconductor contact devices, which typically exhibit ambipolar behavior. Secondly, the maximum attainable $I_d$ values and transconductance increase proportionally with decreasing $L_g$. As $L_g$ decreases from 1 µm to 300 nm, the maximum $I_d$ measured at $V_d = V_g = -2.0$ V increases as 12 µA, 22 µA, and 45 µA, corresponding to the normalized currents of 240 µA·µm$^{-1}$, 420 µA·µm$^{-1}$, and 800 µA·µm$^{-1}$. The $I_d – V_g$ transfer characteristics measured at $V_d$ = -1.0 V show peak transconductance values ($g_m$) of 6.1 µS, 11.5 µS, and 19.6 µS with decreasing $L_g$ from 1 µm to 300 nm . We note that the gate leakage current is below 10 pA in all measurements.

A main finding of our study is summarized in Figure 3. Here we show the total NW FET resistance ($R_m$), measured at small $V_d$, as a function of the geometric channel length ($L_g$), and at different gate overdrive values $|V_g – V_t|$, from 0.5 V to 2.0 V. $V_t$ represents the NW FET threshold gate voltage at which the inversion charge density in the channel is zero. The hole density per unit length ($p$) in the NW FETs is related to the gate bias via: $p = C_{ox} \cdot |V_g – V_t| \cdot e^{-1}$; $C_{ox}$ is the top-gate capacitance per unit length, $e$ is the electron charge. Figure 3(a) data shows that $R_m$, which is the sum of the channel resistance ($R_{ch}$) and the external source and drain contact resistance ($R_{SD}$), is *linear* as a function of $L_g$ for all $|V_g – V_t|$ values. While this is simply a restatement of Ohm's law, the data indicates that transport is diffusive in the Ge-Si$_x$Ge$_{1-x}$ core-shell NWs at room temperature, and allows us to decouple the channel and external contact resistances. The linear fits to $R_m$ vs $L_g$ data at various $|V_g – V_t|$ values have a common intercept, which represents the external contact resistance $R_{SD}$ and the channel length reduction $\Delta L$, namely the difference between the geometric gate length ($L_g$) and the effective channel length ($L_{eff}$) [22]. Figure 3(a) data correspond to $R_{SD}$ = 12.7 kΩ and $\Delta L$ = 43 nm. We note that $R_{SD}$ represents the sum of NW resistance of the highly doped section not covered by the top-gate and the metal-nanowire contact resistance.

To determine the effective mobility in our NW FETs, the $C_{ox}$ values are first calculated self-consistently using Sentaurus TCAD simulation (Synopsys®). The device structure used in simulations is shown in Figure 3(b) (inset). It consists of a Ge core of varying size, a 4 nm-thick Si$_{0.3}$Ge$_{0.7}$ shell, and with a HfO$_2$ dielectric / TaN metal stack corresponding to the actual device. Applying a negative gate bias initially induces holes in the Ge core, and at sufficiently large gate bias holes start to populate the Si$_{0.3}$Ge$_{0.7}$ shell.

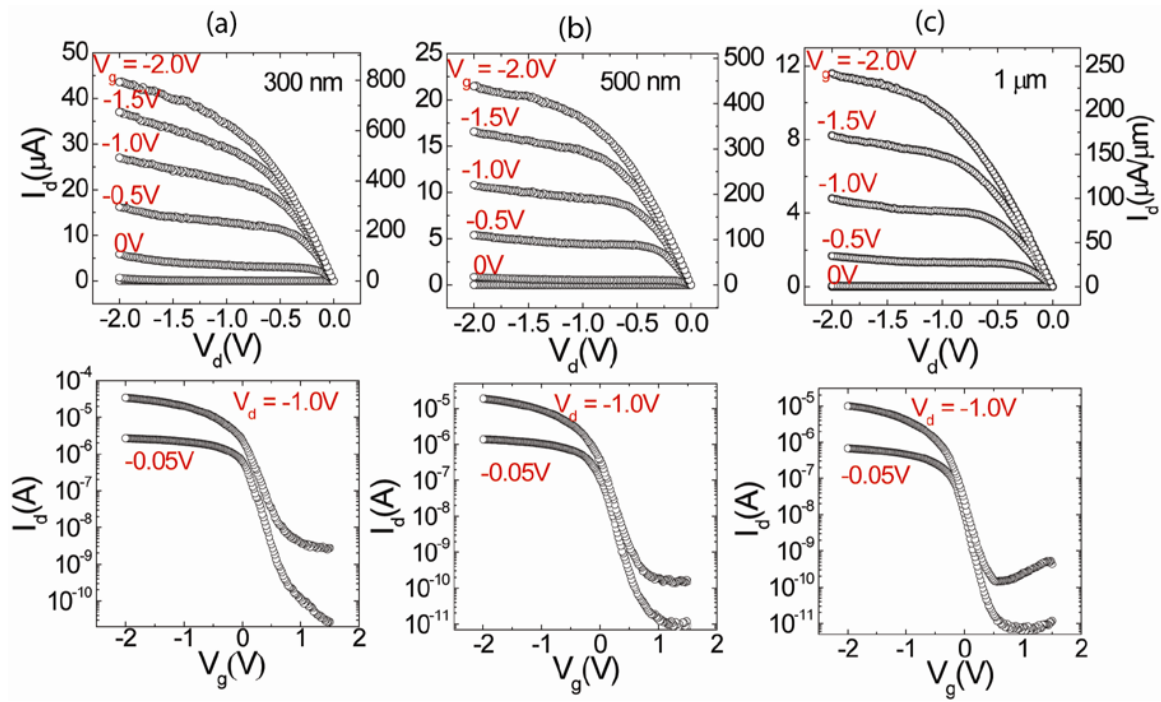

**Figure 2. Electrical characteristics of Ge-Si$_x$Ge$_{1-x}$ core-shell NW FETs at different gate lengths ($L_g$): a.** $L_g$ = 300 nm, $d$ = 55 nm; **b** $L_g$ = 500 nm, $d$ = 49 nm; **c** $L_g$ = 1 µm, $d$ = 48 nm. In each panel the top (bottom) graphs show $I_d$ vs $V_d$ ($I_d$ vs $V_g$) data, measured at constant $V_g$ ($V_d$) values as shown. The right y-axis of the top graphs show $I_d$ normalized to the NW diameter.

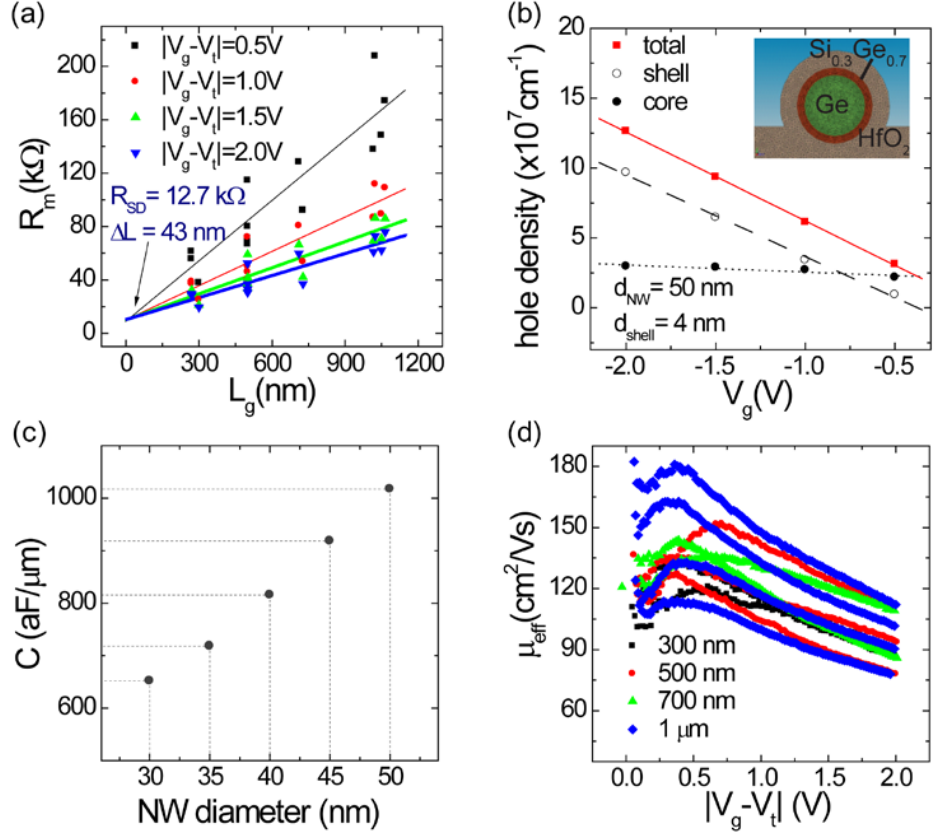

**Figure 3. Channel length resistance scaling and effective mobility extraction a**. Measured device resistance $R_m$ at $V_d$ = -0.05 V vs. $L_g$. The common intercept determines both the source-drain external resistance, $R_{SD}$ = 12.7 kΩ, and the effective channel length reduction $\Delta L$ = 43 nm. **b.** Total hole density vs gate voltage. The total hole density is the sum of the hole density in the NW core and shell. (Inset: schematic representation of the Ω-shape gate of Ge-Si$_x$Ge$_{1-x}$ NW FET) **c.** Capacitance vs. nanowire diameter. The capacitance values were calculated using the relation $C_{ox} = e \cdot (dp/dV_g)$. **d.** Effective mobility of the Ge-Si$_x$Ge$_{1-x}$ core-shell NW FETs for four different channel lengths as a function of gate overdrive.

Figure 3(b) data shows an example of hole densities in the core and shell calculated for a Ge-Si$_{0.3}$Ge$_{0.7}$ core-shell with a 50 nm diameter and a 4 nm – thick shell. We assumed an offset of 0.2 eV between Si$_{0.3}$Ge$_{0.7}$ and Ge valence bands [23-24]. Figure 3c provides the results of the Ω-shape NW FET gate capacitance calculation. The total hole density per unit length in a Ge-Si$_x$Ge$_{1-x}$ NW for a given gate voltage is the sum of the carrier density in the NW core and shell [Figure 3b]. The *p* values are related to $C_{ox}$ and $V_t$ by $e \cdot p = C_{ox} \cdot |V_g - V_t|$; *e* is the electron charge. Thus, total capacitance per unit length is extracted from the equation, $C_{ox} = e \cdot (dp/dV_g)$. Figure 3(c) shows the $C_{ox}$ values calculated for NWs with different diameters. Using the intrinsic channel resistance $R_{ch} = R_m - R_{SD}$ determined from Figure 3(a) data along with the $C_{ox}$, we then extract the *intrinsic* carrier mobility in the Ge-Si$_x$Ge$_{1-x}$ core-shell NWs. The mobility is calculated using $\mu_{eff} = L_{eff} \cdot [R_{ch} C_{ox} (V_t - V_g - 0.5V_d)]^{-1}$, with $V_d$ = -0.05 V. Figure 3(d) data show $\mu_{eff}$ as a function of $|V_g - V_t|$. The results of Figure 3d reveal that the peak hole mobility ranges from 100 cm$^2$(V·s)$^{-1}$ to 180 cm$^2$(V·s)$^{-1}$, values which are up to three-fold higher than that of the Si *p*-MOSFETs with HfO$_2$ gate dielectric [25].

Two main figures of merit for logic devices are the ON-state and OFF-states currents. The ON-state current ($I_{ON}$) determines the FET switching speed, while $I_{OFF}$ determines the passive power consumed by a logic gate (e.g. an inverter). A high-speed, low-power device should possess high $I_{ON}$ and $I_{ON}/I_{OFF}$. To gauge these performance metrics for our Ge-Si$_x$Ge$_{1-x}$ core-shell NW FETs, we define the ON-state current ($I_{ON}$) as the measured $I_d$ at a gate bias $V_{ON} = V_t + (2/3)V_{dd}$, and the OFF-state current ($I_{OFF}$) as the measured $I_d$ at $V_g = V_{OFF} = V_t – (1/3)V_{dd}$; the drain bias in both cases is $V_d = V_{dd}$ = -1.0 V. To estimate the switching speed in our devices we employ the intrinsic gate delay ($\tau$) defined as $\tau = CV/I$, where $C$ is the gate capacitance, $V = V_{dd}$ = -1.0 V, and $I = I_{ON}$ [26]. Figure 4a data shows the relation between $\tau$ and the $I_{ON}/I_{OFF}$ ratio. Here we define a window of $V_{ON} - V_{OFF} = V_{dd}$ = -1 V along the $V_g$ axis to determine $I_{ON}$ and $I_{OFF}$. This graph illustrates the tradeoff between $I_{ON}/I_{OFF}$ and $\tau$ and it shows that $\tau$ decreases as $L_g$ is scaled down. The $I_{ON}/I_{OFF}$ ratio reaches a maximum of up to 10$^4$, a ten-fold higher value than previous results in Ge-Si core-shell NW FETs [12]. The subthreshold slope ($S$), defined as $S = -[d(logI_d)/dV_g]^{-1}$, for different channel lengths is shown in Figure 4(b). These data show $S$ values ranging from 150 to 190 mV·dec$^{-1}$. The measured $S$ values are higher than the thermal limit of 60 mV·dec$^{-1}$, a finding that may be explained by a finite trap density at the dielectric-semiconductor interface. The $S$ value increase as the $L_g$ is reduced, likely because of short channel effect. Lastly, we note that the device performance can be further improved by optimizing the device fabrication process, namely by reducing non-gated the source and drain region as well as by improving the dielectric quality.

## IV. CONCLUSION

We demonstrate high-performance Ge-Si$_x$Ge$_{1-x}$ core-shell NW FETs with highly doped S/D and systematically investigated their scaling properties. Our data allow us to extract key device parameters, such as intrinsic channel resistance, carrier mobility, effective channel length, and external contact resistance, as well as to benchmark the device switching speed and ON/OFF current ratio.

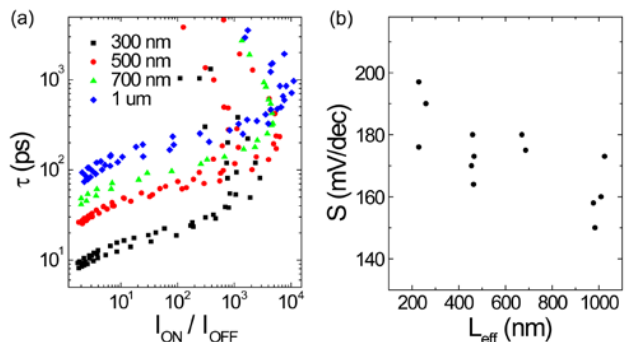

**Figure. 4 Intrinsic gate delay and subthreshold slope a.** Intrinsic gate delay vs. I$_{ON}$/I$_{OFF}$ ratio for different $L_g$ values. **b.** Subthreshold slope vs. effective gate length .

## REFERENCES

[1] S. Datta, T. Ashley, J. Brask, L. Buckle, M. Doczy, M. Emeny, D. Hayes, K. Hilton, R. Jefferies, T. Martin, T. J. Phillips, D. Wallis, P. Wilding, R. Chau, "85nm Gate Length Enhancement and Depletion mode InSb Quantum Well Transistors for Ultra High Speed and very Low Power Digital Logic Applications", *IEDM Tech. Dig.* 2005, pp. 783-786.
[2] S. –H. Lee, Y. Jung, R. Agarwal, "Highly scalable non-volatile and ultra-low-power phase-change nanowire memory", *Nature Nanotech*. 2007, 2, 626-630.
[3] Y. Hu, J. Xiang, G. Liang, H. Yan, C. M. Lieber, "Sub-100 Nanometer Channel Length Ge/Si Nanowire Transistors with Potential for 2 THz Switching Speed", *Nano Lett*. 2008, 8, 925-930.
[4] M. S. Purewal, Y. Zhang, and P. Kim, "Unusual transport properties in carbon based nanoscale materials: nanotubes and graphene", *phys. stat. sol.* (b) 2006, 243, 3418-3422.
[5] S. J. Kang, C. Kocabas, T. Ozel, M. Shim, N. Pimparkar, M. A. Alam, S. V. Rotkin, J. A. Rogers, "High-performance electronics using dense, perfectly aligned arrays of single-walled carbon nanotubes", *Nature Nanotech.* 2007, 2, 230-236.
[6] C. P. Auth, J. D. Plummer, "Scaling theory for cylindrical, fully depleted, surrounding-gate MOSFETs", *IEEE Electron. Dev. Lett*. 1997, 18, 74-76.
[7] N. Singh, A. Agarwal, L. K. Bera, T. Y. Liow, R. Yang, S. C. Rustagi, C. H. Tung, R. Kumar, G. Q. Lo, N. Balasubramanian, D. –L. Kwong, "High-Performance Fully Depleted Silicon Nanowire (Diameter ≤ 5 nm) Gate-All-Around CMOS Devices", *IEEE Electron Dev. Lett*. 2006, 27, 383-386.
[8] M. T. Björk, O. Hayden, H. Schmid, H. Riel, and W. Riess, "Vertical surround-gated silicon nanowire impact ionization field-effect transistors", *Appl. Phys. Lett*., 2007, 90, 142110.
[9] Y. Jiang, T. Y. Liow, N. Singh, L. H. Tan, G. Q. Lo, D. S. H. Chan, D. –L. Kwong, "Performance Breakthrough in 8 nm Gate Length Gate-All-Around Nanowire Transistors using Metallic Nanowire Contacts", *VLSI Tech. Symp.* 2008, 34-35.
[10] D. Wang, Q. Wang, A. Javey, R. Tu, H. Dai, H. Kim, P. C. Mclntyre, T. Krishnamohan, K. C. Saraswat, "Germanium nanowire field-effect transistors with SiO$_2$ and high-κ HfO$_2$ gate dielectrics", *Appl. Phys. Lett*. 2003, 83, 2432-2434.
[11] W. Lu, J. Xiang, B. P. Timko, Y. Mu, and C. M. Lieber, "One-dimensional hole gas in germanium/silicon nanowire heterostructures", *Proc. Natl. Acad. Sci.* U. S. A. 2005, 102, 10046-10051.
[12] J. Xiang, W. Lu, Y. Hu, Y. Wu, H. Yan, C. M. Lieber, " Ge/Si nanowire heterostructures aa high-performance field-effect transistors", *Nature* 2006, 441, 489-492.
[13] W. M. Weber, L. Geelhaar, A. P. Graham, E. Unger, G. S. Duesberg, M. Liebau, W. Pamler, C. Cheze, H. Riechert, P. Lugli, F. Kreupl, "Silicon-Nanowire Transistors with Intruded Nikel-Silicide Contacts", *Nano Lett*. 2006, 6, 2660-2666.
[14] S. –M. Koo, M. D. Edelstein, Q. Li, C. A. Richter, E. M. Vogel, "Silicon nanowire as enhancement-mode Schottky barrier field-effect transistors", *Nanotechnology* 2005, 16, 1482-1485.
[15] Y. Wang, T. –T. Ho, S. Dilts, K. –K. Lew, B. Liu, S. Mohney, J. Redwing, T. Mayer, "Inversion-mode Operation of Thermally oxidized Modulation-Doped Silicon Nanowire Field Effect Devices", *IEEE Dev. Res. Conf. Tech. Dig.* 2006, 175-176.
[16] J. C. Ho, R. Yerushalmi, Z. A. Jacobson, Z. Fan, R. L. Alley, A. Javey, "Controlled nanoscale doping of semiconductors via molecular monolayers", *Nature Materials* 2007, 7, 62-67.
[17] G. M. Cohen, M. J. Rooks, J. O. Chu, S. E. Laux, P. M. Solomon, J. A. Ott, R. J. Miller, W. Haensch, "Nanowire metal-oxide-semiconductro field effect transistor with doped epitaxial contact fro source and drain", *Appl. Phys. Lett*. 2007, 90, 233110.
[18] O. Hayden, M. T. Björk, H. Schmid, H. Riel, U. Drechsler, S. F. Karg, E. Lörtscher, W. Riess, "Fully Depleted Nanowire Field-Effect Transistor in Inversion Mode", *Small* 2007, 3, 230-234.
[19] J. Nah, E. –S. Liu, D. Shahrjerdi, K. M. Varahramyan, S. K. Banerjee, E. Tutuc, "Doping of Ge-Si$_x$Ge$_{1-x}$ core-shell nanowires using low energy ion implantation", *Appl. Phys. Lett*. 2009, 94, 063117.
[20] J. Nah, K. Varahramyan, E. –S. Liu, S. K. Banerjee, E. Tutuc, "Realization of dual-gated Ge-Si$_x$Ge$_{1-x}$ core-shell nanowire field effect transistors with highly doped source and drain", *Appl. Phys. Lett*. 2008, 93, 203108.
[21] A. Satta, E. Simoen, T. Clarysse, T. Janssens, A. Benedetti, B. D. Jaeger, M. Meuris, W. Vandervorst, "Diffusion, activation, and recrystallization of boron implanted in preamorphized and crystalline germanium", *Appl. Phys. Lett*. 2005, 87, 172109.
[22] J. G. J. Chern, P. Chang, R. F. Motta, N. Godinho, "A new method to determine MOSFET channel length", *IEEE Electron Dev. Lett*. 1980, 1, 170-173.
[23] M. M. Rieger, P. Vogl, "Electronic hand parameters in strained Si$_{1-x}$Ge$_x$ alloys on Si$_{1-y}$Ge$_y$ substrates," *Phys. Rev. B* 1993, 48(19), 14276-14287.
[24] F. Schäffler, "High-mobility Si and Ge structures", *Semicond. Sci. Technol*. 1997, 12, 1515-1549.
[25] R. Chau, S. Datta, M. Doczy, B. Doyle, J. Kavalieros, M. Metz, "High-κ/metal-gate stack and its MOSFET characteristics", *IEEE Electron Dev. Lett*. 2004, 25, 408-410.
[26] R. Chau, S. Datta, M. Doczy, B. Doyle, B. Jin, J. Kavalieros, A. Majumdar, M. Metz, M. Radosavljevic, "Benchmarking Nanotechnology for High-Performance and Low-Power Logic Transistor Application", *IEEE. Trans. Nanotechnol*. 2005, 4, 153-158.